\documentclass{aa}
\usepackage{graphics,rotate}

\def\etal{{\it et~al.}}
\def\bec{\begin{center}}
\def\enc{\end{center}}

\def\Mesz{M\'esz\'aros~}
\def\Pacz{Paczy\'nski~}

\def\simg{\mathrel{\hbox{\rlap{\lower.55ex \hbox {$\sim$}}
                   \kern-.3em \raise.4ex \hbox{$>$}}}}
\def\siml{\mathrel{\hbox{\rlap{\lower.55ex \hbox {$\sim$}}
                   \kern-.3em \raise.4ex \hbox{$<$}}}}
\def\Omj{\Omega_j}

\def\eps{\epsilon}
\def\beq{\begin{equation}}
\def\enq{\end{equation}}

\begin{document}

   \thesaurus{13    
              (02.19.1;  
               13.07.1)} 
\title{Gamma Ray Burst Afterglows and their Implications}

\author{P. \Mesz}


\institute{Department of Astronomy \& Astrophysics,
           Pennsylvania State University, University Park, PA 16802, USA\\
              email: nnp@astro.psu.edu}

\date{Received December 15, 1998; accepted -- (astro-ph/9812478)  }

\maketitle

\begin{abstract}

The discovery of X-ray, optical and radio afterglows of GRBs
provides an important tool for understanding these sources. Most
current models envisage GRB as arising in a cataclysmic stellar 
event leading to a relativistically expanding fireball, where particle
acceleration at shocks lead to nonthermal radiation. The predictions of this
scenario are in substantial agreement with the bulk of the observations. 
In addition, the data show a diversity of finer structure behavior, which 
is providing constraints for more detailed models. Current issues of
interest are the implications of the beaming for the energetics, the 
afterglow time structure, dependence on progenitor system, and the role of 
the environment.


\end{abstract}

\section{Introduction: Simple ``Standard" Afterglows}

One can understand the dynamics of the afterglows of GRB in a fairly simple 
manner, independently of any uncertainties about the progenitor systems,
using a relativistic generalization of the method used to model supernova 
remnants. The simplest hypothesis is that the afterglow 
is due to a relativistic expanding blast wave, which decelerates as time goes 
on (\Mesz \& Rees 1997a; earlier simplified discussions were given by 
Katz 1994b, \Pacz \& Rhoads 1993, Rees \& \Mesz 1992).  The
complex time structure of some bursts suggests that the central
trigger may continue for up to 100 seconds. However, at much later
times all memory of the initial time structure would be lost:
essentially all that matters is how much energy and momentum has been
injected; the injection can be regarded as instantaneous in the
context of the much longer afterglow. Detailed calculations and predictions 
from such a model (\cite{mr97a}) preceded the observations of the first 
afterglow detected, GRB970228 (\cite{cos97,jvp97}).

The simplest spherical afterglow model produces a three-segment power law
spectrum with two breaks. At low frequencies there is a steeply rising 
synchrotron self-absorbed spectrum up to a self-absorption break $\nu_a$, 
followed by a +1/3 energy index spectrum up to the synchrotron break $\nu_m$
corresponding to the minimum energy $\gamma_m$ of the power-law accelerated 
electrons, and then a $-(p-1)/2$ energy spectrum above this break, 
for electrons in the adiabatic regime (where $\gamma^{-p}$ is the electron 
energy distribution above $\gamma_m$). A fourth segment and a third break is 
expected at energies where the electron cooling time becomes short compared 
to the expansion time, with a spectral slope $-p/2$ above that. With 
this third ``cooling" break $\nu_b$, first calculated in \cite{mrw98} and 
more explicitly detailed in \cite{spn98}, one has what has come to be called 
the simple ``standard" model of GRB afterglows. This assumes spherical symmetry 
(also valid for a jet whose opening angle $\theta_j \simg \Gamma^{-1}$).
As the remnant expands the photon spectrum moves to lower frequencies, and 
the flux in a given band decays as a power law in time, whose index can change 
as breaks move through it.
\begin{figure}
   \resizebox{\hsize}{!}{\includegraphics{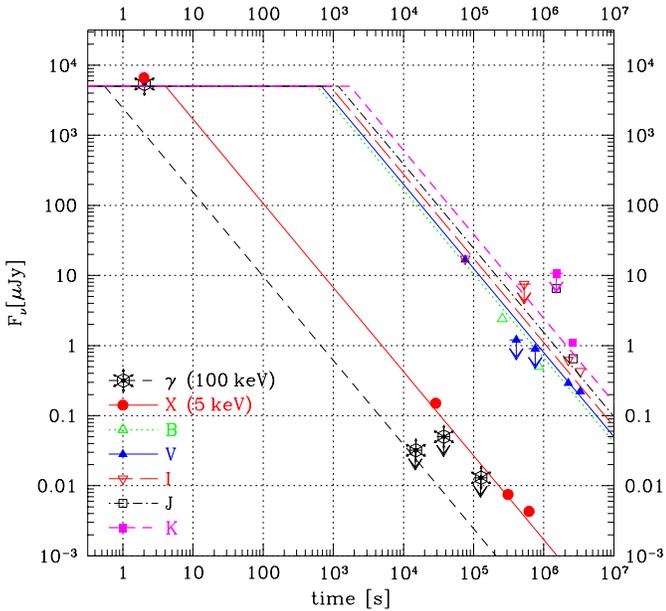}}
   \caption{Light-curves of GRB 970228, compared to the blast wave model 
    predictions of \Mesz \& Rees 1998a (from Wijers et.al 1997)}
   \label{Figure 1}
\end{figure}

The standard model assumes an impulsive energy input lasting much 
less than the observed $\gamma$-ray pulse, characterized by a 
single energy and bulk Lorentz factor value (delta or top-hat function). 
Estimates for the time needed for the expansion to become non-relativistic 
could then be $\siml$ month (\cite{vie97a}), especially if there is an 
initial radiative regime $\Gamma\propto r^{-3}$.
However, even when electron radiative times are shorter than the expansion 
time, it is unclear whether such a regime occurs, as it would
require strong electron-proton coupling (\Mesz, Rees \& Wijers 1998).
The standard spherical model can be straightforwardly generalized
to the case where the energy is assumed to be channeled initially into a 
solid angle $\Omj < 4\pi$. In this case (Rhoads 1997a, 1997b) there is a 
faster decay of $\Gamma$ after sideways expansion sets in, and a decrease 
in the brightness is expected after the edges of the jet become visible, 
when $\Gamma$ drops below $\Omj^{-1/2}$. A calculation using the usual 
scaling laws for a single central line of sight leads then to a steepening of 
the light curve.

The simple standard model has been remarkably successful at explaining 
the gross features of GRB 970228, GRB 970508, etc. (Wijers, Rees \& \Mesz 1997, 
Tavani 1997, Waxman 1997, Reichart 1997). 
Spectra at different wavebands and times have been
extrapolated according to the simple standard model time dependence to get 
spectral snapshots at a fixed time (Waxman 1997, Wijers \& Galama 1998),
allowing fits for the different physical parameters of the burst and 
environment, e.g. the total energy $E$, the magnetic and electron-proton 
coupling parameters ${\eps}_B$ and ${\eps}_e$ and the external density $n_o$. 
In GRB 971214 (Ramaprakash et.al 1997), a similar analysis and the lack of a 
break in the late light curve of GRB 971214 could be interpreted as indicating
that the burst (including its early gamma-ray stage) was isotropic, leading 
to an (isotropic) energy estimate of $10^{53.5}$ ergs. Such large energy 
outputs, whether beamed or not, are quite possible in {\it either} NS-NS, NS-BH 
mergers (\Mesz \& Rees 1997b) or in hypernova/collapsar models (\Pacz 1998,
Popham \etal 1998),
using MHD extraction of the spin energy of a disrupted torus and/or a central
fast spinning BH. However, it is worth stressing that what these snapshot fits 
constrain is only the {\it energy per solid angle} (\Mesz, Rees \& Wijers
1998b).  The expectation of a break after only some weeks or months (e.g., due 
to $\Gamma$ dropping either below a few, or below $\Omega_j^{-1/2}$) is based 
upon the simple impulsive (angle-independent delta or top-hat function) energy 
input approximation. The latter is useful, but departures from it would be 
quite natural, and certainly not surprising.  As discussed below, it would be 
premature to conclude at present that there are any significant constraints 
on the anisotropy of the outflow.
\begin{figure}
   \resizebox{\hsize}{!}{\includegraphics{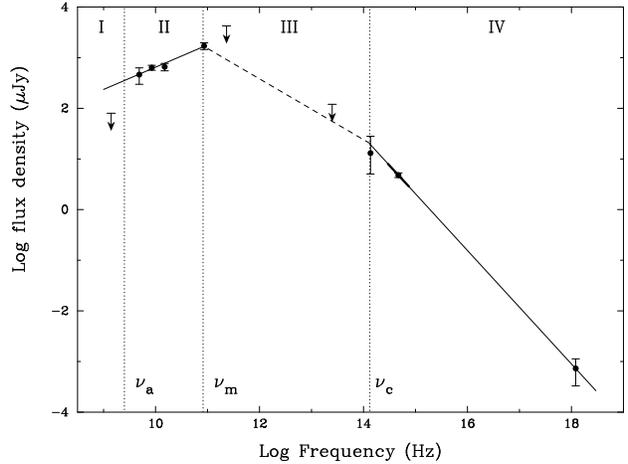}}
  \vskip -2.in
   \caption{Snapshot spectrum of GRB 970508 at $t=12$ days and
 standard afterglow model fit (after Wijers \& Galama 1998)}
   \label{Figure 2}
\end{figure}

\section{``Post-standard" Afterglows Models}

In a realistic situation, one could expect any of several fairly natural
departures from the simple standard model to occur. The first one is that 
departures from a delta top-hat approximation (e.g. having more energy 
emitted with lower Lorentz factors at later times, still shorter than the 
gamma-ray pulse duration) would drastically extend the afterglow lifetime in 
the relativistic regime, by providing a late ``energy refreshment" to the
blast wave on time scales comparable to the afterglow time scale (Rees \&
\Mesz 1998).  The transition to the $\Gamma < \theta_j^{-1}$ regime
occurring at $\Gamma\sim$ few could then occur as late as six months to
more than a year after the outburst, depending on details of the brief
energy input.  

Another important effect is that the emitting region seen by the observer
resembles a ring (Waxman 1997b, Panaitescu \& \Mesz 1998b, Sari 1998).
A numerical integration over angles (Panaitescu \& \Mesz 1998d) shows that 
the sideways expansion effects are not so drastic as inferred from the 
scaling laws for the material along the central-angle line of sight. This 
is because even though the flux from the head-on part of the remnant 
decreases faster, this is more than compensated by the increased emission 
measure from sweeping up external matter over a larger angle, and by the 
fact that the extra radiation, arising at larger angles, arrives later and 
re-fills the steeper light curve. Thus, the sideways expansion (even for a 
simple impulsive injection) actually mitigates the flux decay, rather than 
accelerating it. Combined with the possibility of an extended relativistic 
phase due to nonuniform injection, and the fact that numerical angle 
integrations show that 
any steepening would occur over factors $\sim 2-3$ in time, one must conclude
that we {\it do not} yet have significant evidence for whether the outflow is 
jet-like or not. 
\begin{figure*}
\vskip -0.75in 
\hskip .75in   \resizebox{\hsize}{!}{\includegraphics{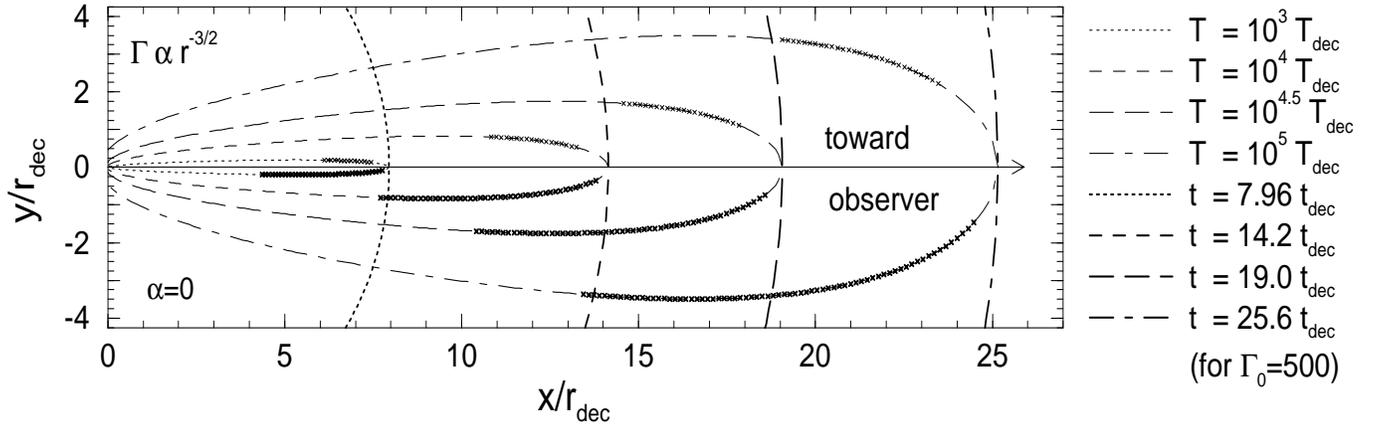}}
\vskip -2.5in
   \caption{Ring-like equal-arrival time $T$ surfaces of an afterglow, based 
   on Panaitescu \& \Mesz 1998d}
   \label{Figure 4}
\end{figure*}

One expects afterglows to show a significant amount of diversity. This
is expected both because of a possible spread in the total energies (or 
energies per solid angle as seen by a given observer), a possible spread 
or changes in the injected bulk Lorentz factors, and also from the 
fact that GRB may be going off in very different environments.
The angular dependence of the outflow, and the radial dependence of the
density of the external environment can have a marked effect on the time
dependence of the observable afterglow quantities (\cite{mrw98}).
So do any changes of the bulk Lorentz factor and energy output during even
a brief energy release episode (\cite{rm98}). 
\begin{figure}
\vskip .2 in
\hskip -.5in   \resizebox{\hsize}{!}{\includegraphics{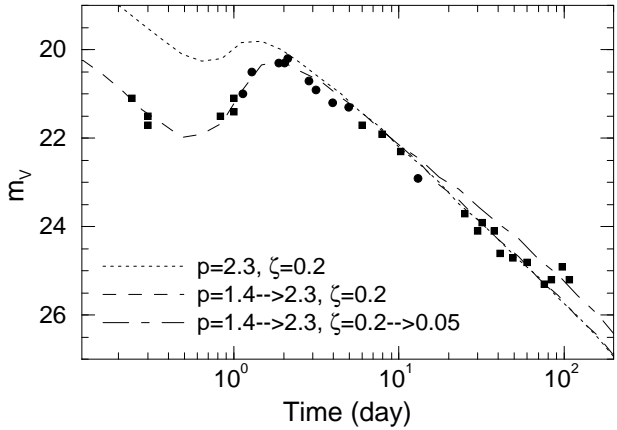}}
   \caption{Optical light-curve of GRB 970508, fitted with a non-uniform
  injection model (Panaitescu, \Mesz \& Rees 1998)}
   \label{Figure 3}
\end{figure}

Strong evidence for departures from the simple standard model is provided by,
e.g., sharp rises or humps in the light curves followed by a renewed decay, 
as in GRB 970508 (\cite{ped98,pir98a}). Detailed time-dependent model fits 
(Panaitescu, \Mesz \& Rees 1998) to the X-ray, optical and radio light curves 
of GRB 970228 and GRB 970508 show that,
in order to explain the humps, a {\it non-uniform} injection or an 
{\it anisotropic} outflow is required. These fits indicate that 
the shock physics may be a function of the shock strength (e.g. the electron
index $p$, injection fraction $\zeta$ and/or $\epsilon_b,~\epsilon_e$ change 
in time), and also indicate that dust absorption is needed to simultaneously 
fit the X-ray and optical fluxes. The effects of beaming (outflow within a 
limited range of solid angles) can be significant (Panaitescu \& \Mesz 1998c), 
but are coupled with other effects, and a careful analysis is needed to
disentangle them. 

Spectral signatures, such as atomic edges and lines, may be expected
both from the outflowing ejecta (\Mesz \& Rees 1998a) and from the external 
medium (Perna \& Loeb 1998, \Mesz \& Rees 1998b, 
Bisnovatyi-Kogan \& Timokhin 1997) in the X-ray and optical 
spectrum of afterglows. These may be used as diagnostics for  the outflow 
Lorentz factor, or as alternative measures of the GRB redshift.
An interesting prediction (\cite{mr98b}; see also \cite{ghi98,bot98}) is that 
the presence of a measurable Fe  K-$\alpha$ {\it emission} line could be a 
diagnostic of a hypernova, since in this case one can expect a massive envelope 
at a radius comparable to a light-day where $\tau_T \siml 1$, capable of 
reprocessing the X-ray continuum by recombination and fluorescence.

The location of the afterglow relative to the host galaxy center can
provide clues both for the nature of the progenitor and for the external 
density encountered by the fireball. A hypernova model would be expected
to occur inside a galaxy, in fact inside a high density ($n_o > 10^3-10^5$).
Some bursts are definitely inside the projected image of the host galaxy, and 
some also show evidence for a dense medium at least in front of the 
afterglow (\cite{ow98}).  On the other hand, for a number of bursts there are 
strong constraints from the lack of a detectable, even faint, host 
galaxy (\cite{sch98}).
In NS-NS mergers one would expect a BH plus debris torus system and 
roughly the same total energy as in a hypernova model, but the mean distance 
traveled from birth is of order several Kpc (\cite{bsp98}),
leading to a burst presumably in a less dense environment. The fits of 
\cite{wiga98} to the observational data on GRB 970508 and 
GRB 971214 in fact suggest external densities in the range of $n_o=$
0.04--0.4 cm$^{-1}$, which would be more typical of a tenuous interstellar 
medium (however, \cite{reila98} report a fit for GRB 980329 with $n_o\sim 10^4$
cm$^{-3}$).  These could arise within the volume of the galaxy, but on
average one would expect as many GRB inside as outside. This is based on
an estimate of the mean NS-NS merger time of $10^8$ years; other estimated 
merger times (e.g. $10^7$ years, \cite{vdh92}) would give a burst much closer 
to the birth site. BH-NS mergers would also occur in timescales $\siml 10^7$ 
years, and would be expected to give bursts well inside the host 
galaxy (\cite{bsp98}).

\section{ Conclusions }
 
The blast wave model of gamma-ray burst afterglows has proved quite 
robust in providing a consistent overall interpretation of the major 
features of these objects at various frequencies. The ``standard model"
of afterglows, involving four spectral slopes and three breaks, is
quite useful in understanding `snaphsot' multiwavelength spectra of
afterglows. However, the constraints on the angle-integrated energy, 
especially at $\gamma$-ray energies, are not strong, and beaming effects 
remain uncertain. Some caution is required in interpreting the observations 
on the basis of the simple standard model. For instance, if one integrates 
the flux over all angles visible to the observer, the contributions from 
different angles lead to a considerable rounding-off of the spectral shoulders, 
so that breaks cannot be easily located unless the spectral sampling is dense 
and continuous, both in frequency and in time. 
Some of the observed light curves with humps, e.g. in GRB 970508, require 
`post-standard' model features (i.e. beyond those assumed in the standard 
model), such as either non-uniform injection episodes or anisotropic outflows. 
Time-dependent multiwavelength fits of this and other bursts also seem to 
indicate that the parameters characterizing the shock physics change with time.
A relatively brief (1-100 s), probably modulated energy input appears the 
likeliest interpretation for most bursts. This can provide an explanation 
both for the highly variable $\gamma$-ray light curves and for
late glitches in the afterglow decays. 

There has been significant progress in understanding how gamma-ray bursts can 
arise in fireballs produced by brief events depositing a large amount of 
energy in a small volume, and in deriving the generic properties of the 
ensuing long wavelength afterglows. There still remain a number of mysteries,
especially concerning the identity of their progenitors, the nature of the 
triggering mechanism, the transport of the energy and the time scales involved. 
However, independently of the details of the gamma-ray burst central engine,
even if beaming reduces their total energy requirements, these objects are 
the most extreme phenomena that we know about in high energy astrophysics, 
and may provide useful beacons for probing the universe at $z \simg 5$. 
With new experiments coming on-line in the near future, there 
is every prospect for continued and vigorous developments both in the 
observational and theoretical understanding of these fascinating objects.
 
\begin{acknowledgements}
I am grateful to Martin Rees for stimulating collaborations
on this subject, as well as to Ralph Wijers, Hara
Papathanassiou and Alin Panaitescu. This research is supported in
part by NASA NAG5-2857
\end{acknowledgements}

\end{document}